\documentclass[aps,prd,amsfonts,amssymb,preprint,eqsecnum,nofootinbib]
{revtex4} 
\usepackage{graphics,epsfig}
\begin{document}
\preprint{WM-05-106}
%
\title{\vspace*{0.5in} Tri-N-ification \vskip
0.1in}
\author{Christopher D. Carone}\email[]{carone@physics.wm.edu}
\affiliation{Particle Theory Group, Department of Physics,
College of William and Mary, Williamsburg, VA 23187-8795}
\date{March 2005}
\begin{abstract}
We consider a natural generalization of trinification to theories
with $3N$ SU(3) gauge groups. These theories have a simple moose 
representation and a gauge boson spectrum that can be interpreted via 
the deconstruction of a 5D theory with unified symmetry broken on a 
boundary.  Although the matter and Higgs sectors of the theory have no 
simple extra-dimensional analog, gauge unification retains features 
characteristic of the 5D theory.  We determine possible assignments of 
the matter and Higgs fields to unified multiplets and present 
theories that are viable alternatives to minimal trinified GUTs.
\end{abstract}
\maketitle

\section{Introduction}\label{sec:intro}
Trinification~\cite{trin1,trin2,trin3,trin4,trin5,trin6} refers to
unified theories based on the gauge group
$G_T=\mbox{SU(3)}_C\times\mbox{SU(3)}_L\times\mbox{SU(3)}_R$, with
gauge coupling equality imposed at a high scale, typically by a
discrete symmetry that cyclically permutes the gauge group labels $C$,
$L$, and $R$.  The Higgs fields responsible for breaking $G_T$ to the
standard model gauge group appear in the ${\bf 27}$-dimensional
representation,
\begin{equation}
\phi({\bf 27}) = \phi_{LR}({\bf 1},{\bf 3},{\bf \bar{3}}) +
\phi_{RC}({\bf \bar{3}},{\bf 1},{\bf 3}) +
\phi_{CL}({\bf 3},{\bf \bar{3}},{\bf 1})  
\,\,\,.
\label{eq:higgs27}
\end{equation}
The component fields $\phi_x$ transform as $({\bf 3}$, ${\bf
\overline{3}})$'s under the pair of gauge groups $x=LR$, $RC$, and
$CL$, respectively.  The theory thus specified has a simple moose
representation, shown in Fig.~\ref{fig1}, with the component fields
$\phi_x$ serving as the `links' that connect neighboring gauge group
`sites'.  This structure is reminiscent of a deconstructed
higher-dimensional theory~\cite{acg,hpw}, aside from the fact that the
link fields in Eq.~(\ref{eq:higgs27}) do not have the same purpose as
in deconstruction, namely, to break chains of replicated gauge groups
down to their diagonal subgroup.

Nevertheless, the structure of Fig.~\ref{fig2} suggests a natural
generalization to trinified theories in which the group factors $C$,
$L$, and $R$ are replicated $N$ times.  We discuss this generalization
in Section~\ref{sec:generalized}.  These theories also have a simple
moose representation, and a gauge spectrum that can be interpreted via
deconstruction.  In particular, we will see that the gauge sector of
the theory is a four-dimensional analog of the 5D trinified theory of
Ref.~\cite{ccjc}, in which gauge symmetry breaking effects are
localized entirely on a boundary.  Such 5D theories have interesting
properties~\cite{nomura}, such as the logarithmic running of the gauge
coupling difference $\alpha_i^{-1}-\alpha_j^{-1}$, for $i\neq j$, and
a delay in the scale of unification above $2\times 10^{16}$~GeV, the
value obtained in the minimal supersymmetric standard model (MSSM).
The construction of purely four-dimensional theories with such
properties is clearly worthy of pursuit, and has lead to interesting
results in the case of SU(5)~\cite{ckt,cmw} and SO(10)~\cite{so10}
unification.  The present work complements this body of literature by
introducing a new class of 4D unified theories that are closely
related to the deconstruction of 5D trinified GUTs.

That said, it will not be the purpose of this paper to present a
literal deconstruction of the theories discussed in Ref.~\cite{ccjc}.
Rather, we proceed from a mostly 4D perspective and develop models
that are viable and economical.  For example, in determining the
embedding of matter fields in the theory, we don't follow the
prescription of deconstruction at all, so that our 4D theory as a
whole cannot be mapped to a local 5D theory in the continuum limit.
The higher-dimensional flavor of gauge unification is nonetheless
retained leading to a new and interesting class of 4D unified
theories.

Our paper is organized as follows.  In Section~\ref{sec:generalized},
we review conventional trinification and define a generalization to
$3N$ replicated SU(3) groups.  In Section~\ref{sec:spectrum}, we study
the gauge boson spectrum of the model, for arbitrary values of the
localized symmetry-breaking vacuum expectation values (vevs). In
Section~\ref{sec:unification}, we apply these results to study gauge
coupling unification in this class of models.  In
Section~\ref{sec:matter}, we describe how one may successfully include
matter and light Higgs fields, so the low-energy particle content is
the same as in the MSSM.  In Section~\ref{sec:conc}, we summarize our
results and suggest directions for future study.

\section{Trinification Generalized}\label{sec:generalized}
\begin{figure}[t]
\epsfxsize 3.3 in \epsfbox{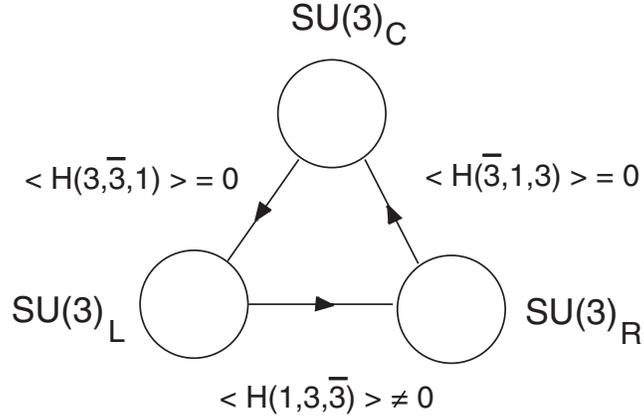} \caption{Moose diagram
for conventional trinified theories.}
\label{fig1}
\end{figure}
Conventional trinified
theories~\cite{trin1,trin2,trin3,trin4,trin5,trin6} are based on the
gauge group $G_T = \mbox{SU(3)}_{C}\times
\mbox{SU(3)}_{L}\times\mbox{SU(3)}_{R}\ltimes Z_3$, where $\ltimes$
indicates a semidirect product.  The $Z_3$ symmetry cyclically
permutes the gauge group labels $C$, $L$ and $R$, ensuring a single
unified coupling at the GUT scale. The breaking of $G_T$ to the
standard model gauge group requires a Higgs field in the ${\bf
27}$-dimensional representation,
\begin{equation}
\phi({\bf 27}) = \phi_{LR}({\bf 1},{\bf 3},{\bf \bar{3}}) +
\phi_{RC}({\bf \bar{3}},{\bf 1},{\bf 3})
+ \phi_{CL}({\bf 3},{\bf \bar{3}},{\bf 1}) 
\,\,\,,
\label{eq:27}
\end{equation}
where the numbers shown represent the SU(3)$_C$, SU(3)$_L$ and SU(3)$_R$
representations, respectively.  At least one such field must develop the
vacuum expectation value,
\begin{eqnarray}
\langle \phi_{LR}\mathbf{(1,3,\bar{3})} \rangle = \left (
\begin{array}{lll}
0 & 0 & 0 \\
0 & 0 & 0 \\
0 & v_2    & v_1
\end{array}
\right ) \,\,\, .
\label{eq:thevevs}
\end{eqnarray}
The presence of vevs only in the third row assures that SU(2)$_L$
remains unbroken, while nonvanishing $32$ and $33$ entries leave
a single unbroken U(1) factor, generated by a linear combination of
the diagonal generators of SU(3)$_L$ and SU(3)$_R$.  Identifying this
with hypercharge $Y$, one finds,
\begin{equation}
A^{\mu}_Y=-\frac{1}{\sqrt{5}}(A^8_L+\sqrt{3}A^3_R+A^8_R)^\mu \,\,\, .
\end{equation}
Note that SU(3)$_L\times$SU(3)$_R$ symmetry allows one to rotate away the
vev $v_2$ in Eq.~(\ref{eq:thevevs}). Therefore, it is usually assumed that
at least two ${\bf 27}$ Higgs fields with vacuum expectation values in
the desired entries are present in the theory.  We will return to this
issue in Sec.~\ref{sec:matter}.  One can verify that this construction 
yields the standard GUT-scale prediction for the weak mixing angle 
$\sin^{2}\theta_{W}=3/8$.

Let us focus on the structure of the gauge and unified-symmetry-breaking
sectors of the theory. A conventional trinified theory can be represented by
the moose diagram shown in Fig.~\ref{fig1}.
For the sake of simplicity we display only  a single Higgs ${\bf 27}$.
The moose representation makes it clear that the ${\bf 27}$ is
anomaly-free, which is relevant for the Higgs representations since
we assume supersymmetry.  The $Z_3$ symmetry is encoded in the
symmetry of the moose under rotations by $120^\circ$.
\begin{figure}[t]
\epsfxsize 3.3 in \epsfbox{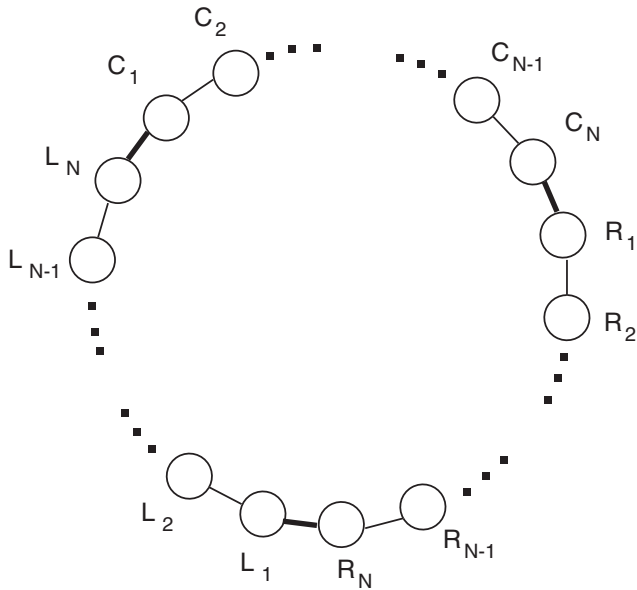} \caption{Generalization to
$3N$ gauge groups (arrows left implicit). The links transforming as
$({\bf \bar{3}}_N,{\bf 1},{\bf 3}_1)$ and
$({\bf 3}_1,{\bf \bar{3}}_N,{\bf 1})$ have no vacuum expectation
values, while the link $({\bf 1},{\bf 3}_1,{\bf \bar{3}}_N)$ has
the vevs given in Eq.~(\ref{eq:thevevs}).}
\label{fig2}
\end{figure}

Now consider the generalization of this moose to $3N$ gauge groups, as
shown in Fig.~\ref{fig2}.  Each link field transforms as a $({\bf
\bar{3}}_i,{\bf 3}_{i+1})$ under consecutive SU(3) groups, reading
around the moose diagram clockwise.  We assume that a generic link has
vev
\begin{equation}
\langle Q_x ({\bf \bar{3}}_j,{\bf 3}_{j+1}) \rangle =
\left(\begin{array}{ccc} v & 0 & 0 \\ 0 & v & 0 \\ 0 & 0 & v
\end{array}\right),
\end{equation}
where $x=C,L$ or $R$. This breaks a chain of SU(3) factors, namely
SU(3)$_{x_i}$ for $i=1\ldots N$, down to its diagonal subgroup. The
diagonal subgroups are precisely
SU(3)$_C\times$SU(3)$_L\times$SU(3)$_R$ of conventional ($N=1$)
trinified theories.  The three links which are not generic are
identified with the links of the $N=1$ theory and have the
corresponding expectation values.  In particular, the links $({\bf
\bar{3}}_N,{\bf 1},{\bf 3}_1)$ and $({\bf 3}_1,{\bf \bar{3}}_N,{\bf
1})$ have no vevs and serve to truncate a linear moose that contains
unbroken SU(3)$_C$. If the link $({\bf 1},{\bf 3}_1,{\bf \bar{3}}_N)$
also had no vev, we could make an analogous statement for SU(3)$_L$
and SU(3)$_R$; we assume, however, that this link has precisely the
expectation value necessary to break SU(3)$_L \times$SU(3)$_R$ down to
the electroweak gauge group of the standard model.  Thus, the total
effect of the link vevs is to break
\begin{equation}
SU(3)^{3N} \rightarrow SU(3)_C \times SU(2)_W \times U(1)_Y,
\label{eq:breaking}
\end{equation}
below the scales $v$, $v_1$ and $v_2$.

At this point, we have said nothing about the values of the $3N$ gauge
couplings.  To maintain gauge unification, we assume that the theory
defined by Fig.~\ref{fig2} is restricted by a $Z_3$ symmetry that sets
equal three sets of $N$ gauge couplings each,
\begin{equation}
g_{C_i} = g_{L_i} = g_{R_i} \,\,\,\,\,\,\,\,\,\, i=1 \ldots N \, .
\end{equation}
Thus, we see that the moose in Fig.~\ref{fig2} is also symmetric under
rotations by $120^\circ$, like the $N=1$ theory.  Unification is
maintained as in the $N=1$ theory since the couplings of the diagonal
subgroups are given by
\begin{equation}
\frac{1}{g_x^2} = \sum_{i=1}^N \frac{1}{g^2_{x_i}}
\end{equation}
for $x=C,L$ or $R$. Note that the vev of the $({\bf 1},{\bf 3}_1,{\bf
\bar{3}}_N)$ link breaks both SU(3)$^{3N}$ and the cyclic symmetry of
the moose.

It is important to note that we are agnostic as to whether the $Z_3$
symmetry is a symmetry of the full theory, including the fermion
representations, or is only an accidental symmetry of the gauge
sector.  The latter could be the case if, for example, all of the
gauge couplings are equal at a high scale due to the dynamics of a
more complete, high-energy theory.   We will remain open to both 
possibilities in our subsequent model building.

In the theory we have described thus far, the special links between
$L_1$-$R_N$, $C_1$-$L_N$ and $R_1$-$C_N$ contribute to the low-energy particle
content of the theory.  On the other hand, we wish only to have two
light Higgs doublets together with the matter content of the MSSM.
How we may arrange for this is discussed in Sec.~\ref{sec:matter}.  We
first, however, address the less model-dependent issue of gauge
unification, assuming that the low-energy matter and Higgs content is
that of the MSSM.

\section{Gauge Boson Spectrum}\label{sec:spectrum}
\begin{figure}[t]
\epsfxsize 5 in \epsfbox{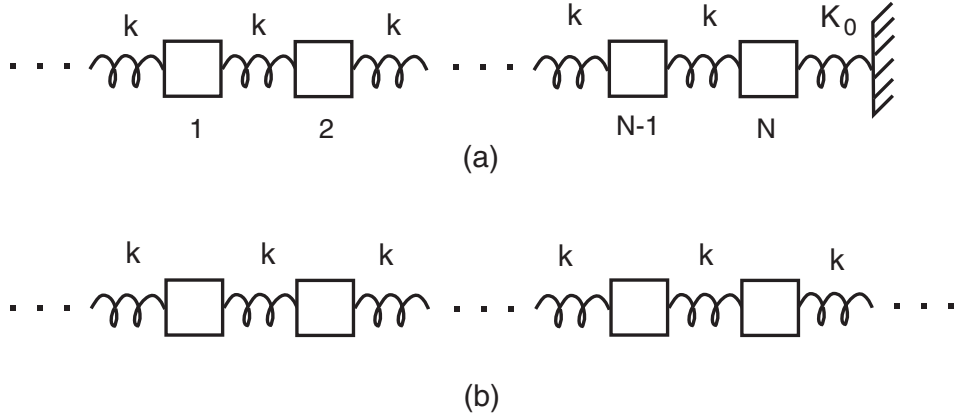} \caption{(a) Mechanical analog 
for studying the gauge boson mass matrix. (b) Translationally invariant
system.}
\label{fig3}
\end{figure}

Let us begin by discussing the spectrum of the $SU(3)_{C_i}$ gauge
multiplets, for $i=1\dots N$.  This portion of the theory is a linear
moose with $N$ unbroken SU(3) factors.  The contribution to the gauge boson
mass matrix from a link spanning the $j$ and $(j+1)^{{\rm th}}$ site
is given by
\begin{equation}
{\cal L}_{j,j+1} = v^2 g^2 \mbox{Tr} [A_C^j A_C^j - 2 A_C^j A_C^{j+1}
+ A_C^{j+1} A_C^{j+1}],
\end{equation}
(with $A_C \equiv A_C^a T^a$) leading to the $N\times N$ mass squared matrix
\begin{equation}
M^2_C = v^2 g^2 \left(\begin{array}{ccccc}
 1 & -1 &    &  &   \\
-1 &  2 & -1 &  &   \\
   &    &  \ddots  & & \\
   &    &  -1   &  2  & - 1\\
   &    &       & -1  & 1 \end{array}\right)
\label{eq:cspec}
\end{equation}
Here we take all SU(3) gauge couplings to be equal, and adopt this
simplifying assumption henceforth. The mass spectrum given by
Eq.~(\ref{eq:cspec}) is that of a deconstructed 5D SU(3) gauge theory,
and is well known~\cite{ckt},
\begin{equation}
m_c = \frac{2}{a} \sin\frac{\ell\pi}{2N} \,\,\,\,\,\,
\,\,\,\, \ell=0,1,2,\ldots \, ,
\label{eq:mspecSM}
\end{equation}
where $a =(vg)^{-1}$ is the lattice spacing.  The $L$ and $R$ sectors of the
moose are more interesting due to the presence of the nontrivial vev at
the $L_1$-$R_N$ link.  Let us define the $i^{\mbox{th}}$ gauge field
as the $16$-component column vector
\begin{equation}
A_i = [A^1_{L_j},\ldots A^8_{L_j},A^1_{R_i},\ldots A^8_{R_i}]  \,\,\,,
\,\,\,\,\, j=N+1-i \,\,\,, \,\,\,\,\, i=1\ldots N\,.
\label{eq:basis}
\end{equation}
If one were to excise the $LR$ sector of the circular moose in
Fig.~\ref{fig2}, and to bend it about the special link at $L_1$-$R_N$,
one would obtain a linear moose with $N$ sites, corresponding to gauge
fields $A_i$ in Eq.~(\ref{eq:basis}).  The symmetry breaking effects
of the special link are confined to the $N^{\mbox{th}}$ lattice site.
Thus, we will be able to interpret our result as a deconstructed 5D
theory with bulk SU(3)$_L\times$SU(3)$_R$ gauge symmetry broken at a
boundary.  Since the $L$-$L$ and $R$-$R$ link fields give identical
contributions to the SU(3)$_L$ and SU(3)$_R$ gauge boson mass
matrices, respectively, we may express the mass matrix for the $A_i$
as
\begin{equation}
M^2_{LR} = v^2 g^2 \left(\begin{array}{ccccc}
 1 & -1 &    &  &   \\
-1 &  2 & -1 &  &   \\
   &    &  \ddots  & & \\
   &    &  -1   &  2  & - 1\\
   &    &       & -1  & 1 \end{array}\right)
+
\left(\begin{array}{ccccc}
 0 &  &    &  &   \\
  &   0  &   &  &   \\
   &    &  \ddots  & & \\
   &    &     &  0  &  \\
   &    &       &   &  \Delta \end{array}\right)
\label{eq:lrspec}
\end{equation}
where each entry represents a $16\times 16$ matrix in $LR$ space. Only the
boundary contribution $\Delta$ has a nontrivial structure in this space --
all others are proportional to an implicit identity matrix. The form
of $\Delta$ is precisely that of the $LR$ gauge boson mass squared matrix
in a conventional, $N=1$ trinified theory.  Let $u$ be the $16$-dimensional
unitary matrix which diagonalizes $\Delta$:
\begin{equation}
u^\dagger \Delta u = \Delta_{\mbox{diag}} \,\,\, .
\end{equation}
Then the $(N,N)$ entry of Eq.~(\ref{eq:lrspec}) may be diagonalized without
affecting any of the others by letting $M^2_{LR} \rightarrow
U^\dagger M^2_{LR} U$, where $U$ is the unitary matrix
\begin{equation}
U = \left(\begin{array}{ccccc}
u & & & & \\
  &u& & & \\
  & &u& & \\
  & & &\ddots& \\
  & & & & u \end{array}\right) \,\,\, .
\end{equation}
Since each $16$-dimensional sub-block is now-diagonalized, we end up with
$16$ decoupled mass matrices, corresponding to the eigenvalues $v^2 g^2 \eta$
of $\Delta$:
\begin{equation}
M^2_\eta = v^2 g^2 \left(\begin{array}{ccccc}
 1 & -1 &    &  &   \\
-1 &  2 & -1 &  &   \\
   &    &  \ddots  & & \\
   &    &  -1   &  2  & - 1\\
   &    &       & -1  & 1+\eta \end{array}\right) \,\,\, .
\label{eq:simplerm}
\end{equation}
Four of the eigenvalues of $\Delta$ are zero, corresponding to the
unbroken SU(2)$_W\times$U(1)$_Y$ gauge bosons, while the remaining $12$
are superheavy. For the massless states ($\eta=0$), Eq.~(\ref{eq:simplerm}) 
reduces to Eq.~(\ref{eq:cspec}), as one would expect, and the mass spectrum is
given by Eq.~(\ref{eq:mspecSM}). The massive gauge fields are more
interesting.  In this case, $\eta$ is nonvanishing and is of the order
$v_i^2/v^2 \geq 1$.  We can find the mass spectrum in the $\eta \neq 0$
case by exploiting a mechanical analogy.  Consider the system of masses
$m$ and springs shown in Fig.~\ref{fig3}a.
The equation of motion for the $i^{\mbox{th}}$ block is given by
\begin{equation}
\frac{d^2 x_i}{dt^2} = - K_{ij} x_j
\end{equation}
where $K_{ij}$ has precisely the form of Eq.~(\ref{eq:simplerm}), with
the identifications $v^2 g^2 = k/m$ and $\eta = K_0/k$.  The squared
frequencies of the normal modes of this mechanical system are precisely
the eigenvalues of $K_{ij}$.  Notice that as $K_0$ is made large ({\em i.e.}
the massive gauge bosons of the last site in the moose are decoupled),
the $N^{\mbox{th}}$ block in our spring system effectively becomes a
fixed wall.  We then obtain an $(N-1)\times (N-1)$ mass matrix of the
same form as Eq.~(\ref{eq:simplerm}) with $1+\eta=2$.  This is exactly
what we expect for the massive gauge modes when the $N^{\mbox{th}}$ site
has a smaller gauge symmetry than the other sites in the moose (see 
for example, Ref.~\cite{ckt}.)

The eigenvalues of $K_{ij}$ may be found by considering the translationally
invariant system shown in Fig.~\ref{fig3}b, and imposing boundary conditions
that mimic the dynamics of the first and $N^{\mbox{th}}$ block of the
system of interest.  Let $\psi(a\,i)$ represent the displacement of the
$i^{\mbox{th}}$ block about its equilibrium position, where $a$ is the
inter-block spacing.  The fact that the first block has no spring to the
left is equivalent to the boundary condition
\begin{equation}
\psi(0)=\psi(a)  \,\,\,
\label{eq:bcleft}
\end{equation}
in the translationally invariant system.  On the other hand, the effect
of the spring with larger spring constant $K_0$ to the right of
the $N^{\mbox{th}}$ block is replicated by the condition
\begin{equation}
\psi(a[N+1]) = -(\eta - 1) \, \psi(aN)  \,\,\, .
\label{eq:bcright}
\end{equation}
In other words, in the infinite system, one requires that the
$(N+1)^{\mbox{th}}$ block moves so that the force on the $N^{\mbox{th}}$
block is indistinguishable from that of a stiffer spring connected
to a wall.  We may thus consider a normal mode solution to the
infinite system of the form
\begin{equation}
\psi \propto e^{i\,k\,x} + B\, e^{-i\,k\,x}
\end{equation}
and determine the wave numbers $k$ allowed by these boundary conditions.
We find that $N-1$ values of $k$ are determined by the transcendental
equation
\begin{equation}
\cot(kaN)\cot(ka/2) = -(1-2/\eta)
\label{eq:telight}
\end{equation}
leading to the eigenvalues
\begin{equation}
m^2 = \frac{4}{a^2} \sin^2\left(\frac{ak}{2}\right) \,\,\,.
\label{eq:mslight}
\end{equation}
Here, $m^2$ represents the squared frequencies of the normal mode solutions
in the mechanical system, and  $N-1$ gauge boson squared
masses in the problem of interest.  The $N^{\mbox{th}}$ gauge
boson mass, however, is not given by Eqs.~(\ref{eq:telight}) and
(\ref{eq:mslight}). The reason is that we have assumed that $k$ is
real; complex $k$ is perfectly consistent with the translation
invariance of the infinite system (see, for example, Ref.~\cite{waves}).
Taking $k=k_R + i k_I$, one finds another solution:
\begin{eqnarray}
k_R = n\pi/a \,,\,\,\,\, n &=&\mbox{integer} \nonumber \\
\coth(k_IaN)\tanh(k_Ia/2) &=& (1-2/\eta) \,\,\, ,
\label{eq:teheavy}
\end{eqnarray}
with
\begin{equation}
m^2 = \frac{4}{a^2} \cosh^2\left(\frac{ak_I}{2}\right) \,\,\,.
\label{eq:msheavy}
\end{equation}
It is not hard to verify that  Eqs~(\ref{eq:telight}), (\ref{eq:mslight}),
(\ref{eq:teheavy}) and (\ref{eq:msheavy}) are correct.  For example, a
two-by-two matrix with $a=1$ and $\eta=4$, has eigenvalues
$m^2=3\pm\sqrt{5}$.   The corresponding solutions to Eqs.~(\ref{eq:telight})
and (\ref{eq:teheavy}) are $k=-0.9045569$ and $k_I=-1.0612751$, which
yield precisely the same results via Eqs.~(\ref{eq:mslight}) and
(\ref{eq:msheavy}).

The parameter $\eta$ allows us to interpolate between a number of familiar
limits.  For example, the choice $\eta=1$ yields the mass matrix for
a gauge boson distributed among $N$ sites that receives no mass
contribution from the $(N+1)^{\mbox{th}}$ link field.  This is the case,
for example, for the massive $X$ and $Y$ bosons in an SU(5) moose
with $N+1$ sites, in which the gauge symmetry of the last site is
taken to be SU(3)$_C\times$SU(2)$_W\times$U(1)$_Y$.  In this
case, Eq.~(\ref{eq:telight}) reduces to
\begin{equation}
\cos(ka[N+1/2])=0
\end{equation}
which is solved by
\begin{equation}
ka = \pi (2 n+1) /(2 N+1), \,\,\,\,\, n=0,1,\ldots \,\,\, .
\label{eq:exone}
\end{equation}
This is consistent with the results of Ref.~\cite{ckt}, which presents the
spectrum of a deconstructed 5D SU(5) theory with unified symmetry broken
explicitly at an orbifold fixed point.  Since we generally assume that
$v_i \gg v$, and hence that $\eta \gg 1$, a more relevant limit is the
one in which $\eta \rightarrow \infty$.  As we described earlier, this
corresponds to a physical system in which the $N^{\mbox{th}}$ block has
effectively become fixed.  In this case, Eq.~(\ref{eq:telight}) reduces to
\begin{equation}
\cos(ka[N-1/2])=0
\end{equation}
which is solved by
\begin{equation}
ka = \pi (2 p+1) /(2 N-1), \,\,\,\,\, p=0,1,\ldots \,\,\, .
\label{eq:extwo}
\end{equation}
As we expect, this result is simply Eq.~(\ref{eq:exone}) with
the replacement $N \rightarrow N-1$.  For large values of $N$,
the expression for $m$ is approximately
\begin{equation}
m \approx k \approx \frac{\pi}{2Na} (2 p + 1) \equiv
\frac{M_c}{2} (2p+1) \,\,,\,\,\,\,\,\,\,p=0,1,\ldots
\end{equation}
where we have identified the compactification scale $M_c =
\pi/Na$. This is shifted relative to the gauge bosons with zero modes,
which have the spectrum $m \approx M_c \, p$ in the same limit.  This
is exactly the behavior we expect in the Higgsless
theory~\cite{terning}, obtained when when takes the boundary vevs $v_i
\rightarrow \infty$.  Thus, the present work demonstrates how one may
obtain a four-dimensional analog for the 5D Higgsless trinified model
described in Ref.~\cite{ccjc}.

With Eqs.~(\ref{eq:mspecSM}), (\ref{eq:telight}), (\ref{eq:mslight}),
(\ref{eq:teheavy}), and (\ref{eq:msheavy}) in hand, we have all
the information we need to take into account the effect of finite
boundary vevs on the gauge boson mass spectrum.  We will use this result
in our study of gauge unification in the following section.
\section{Gauge Unification}\label{sec:unification}
\begin{figure}[t]
\epsfxsize 4 in \epsfbox{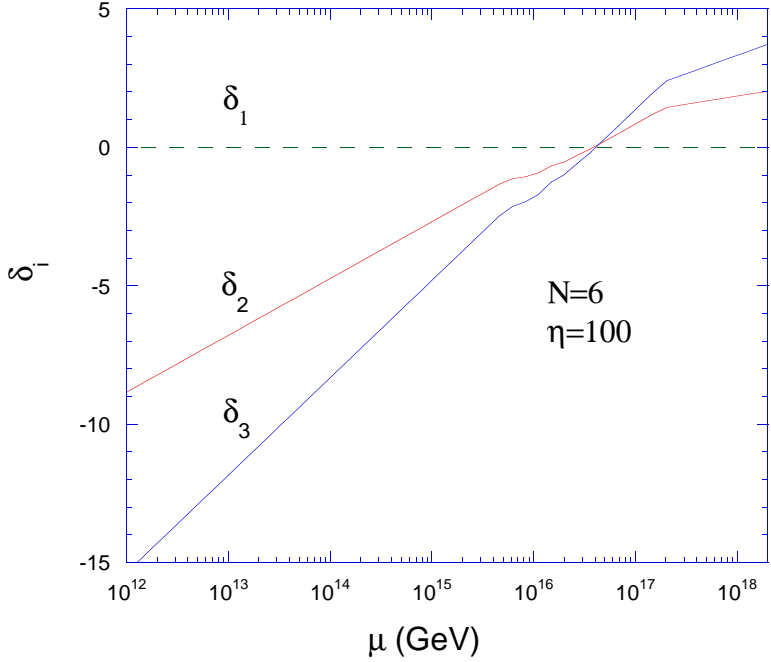} \caption{Gauge unification for $N=6$
and $\eta=100$. See the text for details.}
\label{fig4}
\end{figure}

Aside from the ${\cal N}=2$ vector supermultiplets that we expect at
each massive KK level of the deconstructed theory~\cite{je1,je2}, we assume 
that the only other fields relevant for unification are the light matter and
Higgs fields of the MSSM. We justify this assumption in
Sec.~\ref{sec:matter}, where we demonstrate how this can be arranged.
We will find it convenient to express our results in terms of
the differences
\begin{equation}
\delta_i(\mu) \equiv \alpha^{-1}_i(\mu) - \alpha^{-1}_1(\mu)  \,\,\, ,
\end{equation}
where $\mu$ is the renormalization scale. Since the theory of interest
to us here represents a deconstructed version of the 5D trinified theory
of Ref.~\cite{ccjc}, the basic quantities of interest in studying
gauge unification at the one loop level have the same form:
\begin{equation}
\delta_i(\mu) = \delta_i(m_H^{(1)}) - \frac{1}{2\pi} R_i(\mu) \,\,\,,
\label{eq:above}
\end{equation}
\begin{equation}
R_2(\mu) = -\frac{28}{5} \ln\left(\frac{\mu}{m_H^{(1)}}\right)
-4 \sum_{0<m_0^{(n)}<\mu}\ln\left(\frac{\mu}{m_0^{(n)}}\right)
+4 \sum_{0<m_H^{(n)}<\mu}\ln\left(\frac{\mu}{m_H^{(n)}}\right) \,\,\,,
\end{equation}
\begin{equation}
R_3(\mu) = -\frac{48}{5} \ln\left(\frac{\mu}{m_H^{(1)}}\right)
-6 \sum_{0<m_0^{(n)}<\mu}\ln\left(\frac{\mu}{m_0^{(n)}}\right)
+6 \sum_{0<m_H^{(n)}<\mu}\ln\left(\frac{\mu}{m_H^{(n)}}\right) \,\,\, .
\end{equation}
Here, $m_0^{(n)}$ and $m_H^{(n)}$ represent the mass levels corresponding
to gauge fields with and without zero modes, respectively.  These expressions
are valid for $\mu > m_H^{(1)}$, which we assume to be the lightest gauge
field KK excitation (recall that as $\eta \rightarrow \infty$,
$m_H^{(1)} \rightarrow m_0^{(1)}/2$).  The unification scale $M_{GUT}$
is identified with the scale at which the moose is reduced to the diagonal
subgroup SU(3)$_C\times$SU(3)$_L\times$SU(3)$_R$; this is usually taken to
be $2 a^{-1} = 2 v g$, the scale of the heaviest KK excitation.  Thus
we require
\begin{equation}
\delta_2(2 a^{-1})=0 \,\,\, ,
\label{eq:ucond}
\end{equation}
which determines the unification scale in terms of $3N$ and $\eta$.  For
arbitrary $\eta$, Eq.~(\ref{eq:ucond}) can be solved numerically; for example,
with $N=6$ and $\eta=100$, one finds that $M_{\mbox{GUT}}=2 a^{-1}$ is
$3.85 \times 10^{16}$~GeV, as is shown in Fig.~\ref{fig4}.
The changes in slope of $\delta_2$ and $\delta_3$ above
$m_H^{(1)} \approx 0.284\,a^{-1}$ correspond to vector multiplet mass
thresholds.  The one kink present above the unification scale corresponds
to the mass threshold that decouples in the $\eta \rightarrow
\infty$ limit.  One finds that the quality of unification in this
example is given by
\begin{equation}
\frac{\delta\alpha_3^{-1}(2a^{-1})}{\alpha_1^{-1}(2a^{-1})} \approx -0.3\%.
\label{eq:n6uq}
\end{equation}
Note that the unification scale is above $2\times 10^{16}$~GeV, the unification
scale of conventional supersymmetric theories.

Since we are interested in $\eta\gg 1$, we will analyze the general case as
an expansion in $1/\eta$.  The results we present below may be derived by
taking into account the following simplifications,
\begin{equation}
\sum_{0<m_0^{(n)}<\mu} \ln m_0^{(n)} =
\frac{1}{2}(N-1) \ln\left(\frac{4}{a^2}\right)
+\frac{1}{2} \ln \left(\frac{4N}{2^{2N}}\right) \,\,\, ,
\label{eq:sumzero}
\end{equation}
and
\begin{equation}
\sum_{0<m_H^{(n)}<\mu} \ln m_H^{(n)}= -\frac{1}{2} (N-1) 
\ln a^2 - \frac{1}{2\eta}
+ {\cal O}(1/\eta^2) \,\,\, ,
\label{eq:sumH}
\end{equation}
which are valid if the renormalization scale is above all but the last
mass threshold $m_H^{(N)}$ ({\em i.e.}, the kink above the unification
point shown in Fig.~\ref{fig4} is not included in the sums).  Then it
follows that
\begin{equation}
2 a^{-1} \approx 2\times 10^{16}\mbox{ GeV} \times \, e^{5/(14\eta)} \, 
N^{5/14} \,\,\, ,
\end{equation}
which clarifies the previously noted delay in unification by showing the
scaling in $N$ and the weak dependence on $\eta$, for $\eta$ large.  One
may also use Eqs.~(\ref{eq:sumzero}) and (\ref{eq:sumH}) to obtain
a general expression for the quality of unification ({\em e.g.}
Eq.~(\ref{eq:n6uq})).
One finds,
\begin{equation}
\frac{\delta\alpha_3^{-1}(2a^{-1})}{\alpha_1^{-1}(2a^{-1})} = -0.85\% \cdot
f(N,\eta) \,\,\, ,
\end{equation}
where
\begin{equation}
f(N,\eta) = \frac{1-0.340(\ln N + 1/\eta)}{1+0.028N-0.016\ln N+ 0.004/\eta} 
\,\,.
\end{equation}
The function $f$ ranges approximately from $-0.15$ to $0.97$, for large 
$\eta$, assuring reasonably accurate unification.

\section{Light Higgs and Matter Fields}\label{sec:matter}
\begin{figure}[t]
\epsfxsize 3.5 in \epsfbox{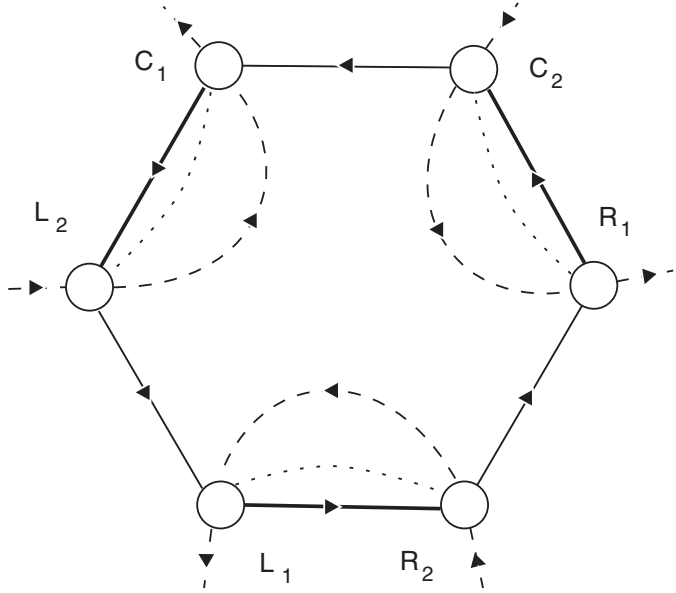} \caption{Embedding of
${\bf \overline{27}}$ (dashed lines) and ${\bf 192}$ (dotted lines)
Higgs fields, which get vevs only at the $L_1$-$R_2$ link. Exterior lines
represented spectators for anomaly cancellation.}
\label{fig5}
\end{figure}

We now show how one may construct tri-N-ified theories with the
low-energy particle content of the MSSM.  Our first example assumes
that the $Z_3$ symmetry applies to the theory as a whole, including
the fermion representations, which makes achieving the desired
low-energy theory nontrivial.  Let us consider a viable model for the
next-to-minimal case of $N=2$.

We would first like to arrange for precisely two Higgs doublet fields
in the low-energy spectrum.  We follow the approach of
Ref.~\cite{ccjc} and consider doublets living within ${\bf 27}$ and
${\bf \overline{27}}$ representations of the diagonal subgroup, and an
additional Higgs field in the ${\bf 192}$-dimensional representation,
to arrange for doublet-triplet splitting.  The Higgs ${\bf 27}$
consists of the `special' links in the original circular moose,
namely, the links between the sites $L_1$-$R_2$, $R_1$-$C_2$, and
$C_1$-$L_2$.  The main purpose of these links was to cancel anomalies,
and to provide for the breaking of the diagonal gauge group at a
boundary.  We add the ${\bf \overline{27}}$ as shown by the dashed
lines in Fig.~\ref{fig5}.  Notice that to avoid anomalies, we have
added spectator fields; for example, the $({\bf 1},{\bf
\bar{3}}_1,{\bf 3}_2)$ component requires three spectators with
quantum numbers $S^a_{L_1} \sim ({\bf 1},{\bf 3}_1,{\bf 1})$, for
$a=1\ldots 3$ and three with quantum numbers $S^a_{R_2}\sim({\bf
1},{\bf 1},{\bf \bar{3}}_2)$.  The links that acquire diagonal vevs
generate mass terms for these spectators,
\begin{equation}
W = \lambda^{ab}_R S^b_{R_1} ({\bf 3}_1) Q_R({\bf \bar{3}}_1,{\bf 3}_2)
S^a_{R_2}({\bf \bar{3}}_2)   + (R \rightarrow C) + (R \rightarrow L) \,\,\, ,
\end{equation}
where the $\lambda$ are coupling constants.  While the spectators are
present at the unification scale, they do not alter the
renormalization group analysis presented in the previous section.
With both ${\bf 27}$ and ${\bf \overline{27}}$ Higgs fields present in
the theory, we can arrange for heavy masses for all the color-triplet
components.  Doublet-triplet splitting may be obtained by
adding the Higgs representation $\Omega({\bf 192}) \sim
({\bf 1},{\bf 8}_1,{\bf 8}_2) + ({\bf 8}_2,1,{\bf 8}_1)
+ ({\bf 8}_1,{\bf 8}_2,1)$.  Notice that the representation
$({\bf 1}, {\bf 8}_i, {\bf 8}_j)$ is anomaly free, so we
do not require any additional spectator fields.  We assume that only
the $({\bf 1},{\bf 8}_1,{\bf 8}_2)$ component of the ${\bf 192}$ and
the $({\bf 1},{\bf \bar{3}}, {\bf 3})$ of the ${\bf \overline{27}}$
acquire GUT-breaking vacuum expectation values, which again localizes
the breaking of the diagonal subgroup to the $L_1$-$R_2$ link. These
new vevs contribute to the the matrix $\Delta$ in
Eq.~(\ref{eq:lrspec}), but otherwise do not alter the discussion of
Section~\ref{sec:unification}.  In addition, appropriate vevs in the
${\bf 192}$ and the ${\bf \overline{27}}$ prevent one from eliminating
the desired symmetry-breaking effect of Eq.~(\ref{eq:thevevs}) via a
gauge transformation. Calling the ${\bf \overline{27}}$ Higgs
$\overline{Q}$, the superpotential terms relevant to doublet-triplet
splitting are given by
\begin{equation}
W = \mu Q_{R_1 C_2} \overline{Q}_{R_1 C_2} + \mu Q_{C_1 L_2} 
\overline{Q}_{C_1 L_2}
+ Q_{L_1 R_2} (\mu + h \Omega_{L_1 R_2} )\overline{Q}_{L_1 R_2} \,\,\, .
\end{equation}
As described in Ref.~\cite{ccjc}, a vev of the form
\begin{equation}
\langle \Omega^{\alpha\beta}_{ab} \rangle = v_\Omega\,
{T^8}^a_b \, {T^3}^\alpha_\beta
\end{equation}
allows one to make a doublet component of $Q_{L_1 R_2}$ and a doublet
component of $\overline{Q}_{L_1 R_2}$ light, provided that $\mu
\approx -4\sqrt{3} h v_\Omega$.  The light components, which we
identify as the two Higgs doublets of the MSSM are localized at the
$L_1$-$R_2$ link.  Note that the down-type Higgs doublet lives within
the ${\bf \overline{27}}$ representation, which does not have a direct
Yukawa coupling with matter fields transforming in the ${\bf
27}$. However, $\overline{Q}^2/\Lambda$ transforms as a ${\bf 27}$,
where $\Lambda$ is an ultraviolet cutoff.  This combination provides
for down-type Higgs Yukawa couplings via higher-dimension operators,
suppressed relative to the up-type couplings by a factor of
$v_i/\Lambda$.

\begin{figure}[t]
\epsfxsize 3.0 in \epsfbox{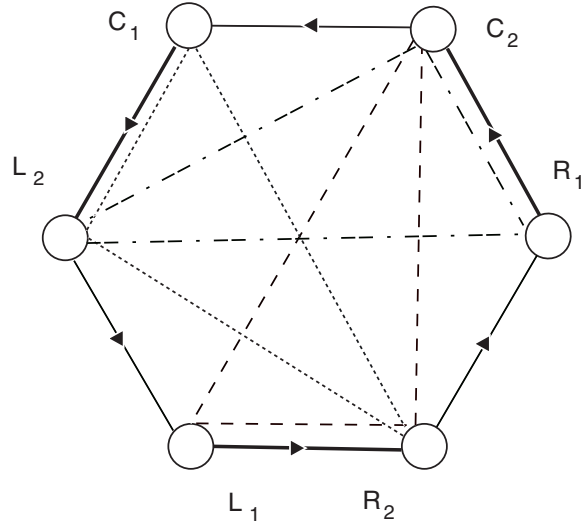} \caption{Placement of three generations
of matter fields in the $N=2$ model. The dotted, dot-dashed and dashed lines
correspond to the first, second and third generations, respectively.}
\label{fig6}
\end{figure}

We now must arrange for the embedding of three generations of matter fields,
so that there are adequate couplings to the light, localized Higgs doublets.
A possible configuration is shown in Fig.~\ref{fig6}.
Here we have chosen to embed the matter fields so that the moose
remains invariant under $120^\circ$ rotations.  One could argue,
therefore, that the $Z_3$ symmetry forces the existence of three
generations in this model. Since the light Higgs doublets are only
present on the $L_1$-$R_2$ link, while the generations are delocalized
we expect that Yukawa couplings originate via higher-dimension
operators, suppressed by powers of $v/\Lambda$.  For the model to be
successful, these suppression factors must not be so great that we
are prevented from achieving phenomenologically viable Yukawa textures.
Let us identify the matter ${\bf 27}$'s by the sites that they connect, {\em
e.g.}, $\psi(212)$ transforms under the subset of gauge groups
($C_2$,$L_1$,$R_2$).  Then we identify the three generations as
follows:
\begin{eqnarray}
\psi^{122} &=& \mbox{generation 1} \nonumber \\
\psi^{221} &=& \mbox{generation 2} \\
\psi^{212} &=& \mbox{generation 3} \,\,\, .\nonumber
\end{eqnarray}
Given this choice, we see, for example, that the third generation
Yukawa couplings are not suppressed,
\begin{equation}
W \supset h_t H_u({\bf 1},{\bf 3}_1,{\bf \overline{3}}_2) \,
\psi_{RC}^{212}({\bf\overline{3}_2},{\bf 1},{\bf 3}_2) \,
\psi_{CL}^{212}({\bf 3}_2,{\bf\overline{3}_1},{\bf 1}) \,\,\, ,
\end{equation}
while a purely second generation coupling
\begin{equation}
\frac{1}{\Lambda^2} h_c H_u({\bf 1},{\bf 3}_1,{\bf \overline{3}}_2) \,
\psi_{RC}^{221}({\bf \overline{3}}_2,{\bf 1},{\bf 3}_1) \,
\psi_{CL}^{221}({\bf 3}_2,{\bf \overline{3}}_2,{\bf 1}) \,
Q_L ({\bf \overline{3}}_1,{\bf 3}_2) \, Q_R({\bf \overline{3}}_1,{\bf 3}_2)
\end{equation}
involves a suppression factor of order $v^2/\Lambda^2$.  One therefore
finds that the entries of the up quark Yukawa matrix involve suppression
factors summarized by
\begin{equation}
\mbox{Gauge suppression factors} \sim
\left(\begin{array}{ccc} \xi & \xi^2 & \xi^2 \\
\xi^3 & \xi^2 & \xi \\ \xi^2 & \xi & 1 \end{array}\right)  \,\,\, ,
\label{eq:supfacts}
\end{equation}
where we have defined $\xi=v/\Lambda$; the down sector suppression
factors are one power of $v_i/\Lambda$ higher.  The tightest constraint
on the size of these parameters comes from the strange quark Yukawa 
coupling, which requires that $\xi \sqrt{v_i/\Lambda} \agt 1/20$. 
Provided that this is satisfied, one may choose coupling matrices $h_u$ 
and $h_d$ to obtain the additional flavor suppression required to produce 
Yukawa textures that are phenomenologically viable (for example, compare 
to those found in Ref.~\cite{u2}). The reader can check that the same is 
true in the charged lepton sector as well.

In addition, the ${\bf 27}$'s of matter fields contain exotic particle
content that becomes heavy in minimal trinified theories due to the
SU(3)$^3$-breaking vevs of the $L_1$-$R_2$ link.  The same happens
here, though there is some suppression due to the delocalization of
the matter multiplets. Since a generic link field transforms as $Q
\sim ({\bf \bar{3}}_i,{\bf 3}_{i+1})$, and it is possible to arrange
$Q^2/\Lambda \sim ({\bf 3}_i,{\bf \bar{3}}_{i+1})$, by elementary
SU(3) group theory, one has the necessary building blocks to
reproduce {\em any} mass operator of the $N=1$ theory at some order in
$\langle Q \rangle/\Lambda$.  Note that the exotic fields in the ${\bf
27}$ form complete SU(5) multiplets~\cite{ccjc}, so they can appear
below the unification scale without substantially altering the
conclusions of Section~\ref{sec:unification}.

While the previous model can accommodate the flavor structure of
the standard model, it is not clear whether successful models can
be constructed for larger $N$.  The assumption of the unbroken
$Z_3$ symmetry would force different generations to be widely
separated if the number of sites is large.  Let us consider the
possibility that the unified boundary condition on the gauge couplings
has a dynamical origin and is not due to symmetry. Equivalently, we may
assume that the $Z_3$ symmetry is an accidental symmetry of the gauge and
Higgs sector, but not of the matter fields.  Then we may place all three
generations on a single set of gauge sites, namely, $C_j$-$L_1$-$R_2$, for
some $j$.  In this case, no suppression factors appear in the Yukawa
couplings, and the origin of the fermion mass hierarchy is completely
decoupled from the physics of unification.  While we keep the arrangement
of Higgs fields the same, we can add an additional three pairs of
${\bf 3}$ and ${\bf \overline{3}}$ spectator fields at each new
gauge site; in this way, all spectators will acquire GUT-scale masses 
by coupling in pairs via an intermediate link field.  For either the
$C$, $L$ or $R$-sector spectators, these interactions have the form
\begin{equation}
W = \sum_{i=1}^{N-1}
\lambda^{ab}_i S^a_i({\bf 3}_i) \,
Q({\bf \overline{3}}_i, {\bf 3}_{i+1})\, S^b_{i+1}({\bf
\overline{3}}_{i+1})  \,\,\, .
\end{equation}
Using this construction, we can study models with many more gauge
sites. The only limitation on the size of $N$ comes from the assumption
that the SU(3) gauge couplings $g$ remain perturbative.  For the case
of identical gauge couplings, $g$ is related to the gauge coupling
of the diagonal subgroup, $g_{\mbox{{\tiny GUT}}}$, by
\begin{equation}
g = \sqrt{N} g_{\mbox{{\tiny GUT}}} \,\,\, ,
\end{equation}
Using our earlier renormalization group analysis, one finds numerically 
that the constraint $g^2/(4\pi) < 1$ translates to
\begin{equation}
N \leq 65 \,\,\, .
\end{equation}
\section{Conclusions}\label{sec:conc}

We have considered a natural generalization of trinification to theories
with $3N$ gauge groups.  We have showed that the gauge and Higgs sector
of the theory can be interpreted as a deconstructed version of a 5D
trinified theory with unified symmetry broken at a boundary; we have
studied the gauge spectrum in the deconstructed theory for arbitrary
boundary Higgs vevs. As in the 5D theory, the differences in inverse gauge
couplings evolve logarithmically, and one finds successful unification at
a scale higher than in the 4D MSSM.  Since the unification scale grows
as $N^{5/14}$, unification may be delayed to $9 \times 10^{16}$~GeV
before the gauge coupling of the replicated SU(3) factors becomes
nonperturbative.   This might be compared to the result in the 5D theory,
$1 \times 10^{17}$~GeV, which is the point at which the 5D Planck scale and
the unification scale coincide~\cite{ccjc}.

We have also considered ways in which the light matter and Higgs
content of the MSSM could be embedded in the (rather large) unified
group.  We first showed how doublet-triplet splitting could be
arranged, via fine tuning, leaving light Higgs doublets localized
between two gauge sites in the original moose.  The point here is not
to solve the doublet-triplet splitting problem, but to present a
viable zeroth-order framework in which such questions can be studied.
With all symmetry breaking localized, we then considered the embedding
of matter fields.  Assuming first that the unified boundary condition
on the gauge couplings arises from a $Z_3$ symmetry that rotates the
moose by $120^\circ$, one obtains symmetrical embeddings of the
fermion multiplets within the moose ({\em c.f.}, Fig.~\ref{fig6}).
Although these have no strict extra-dimensional interpretation, we
have, roughly speaking, delocalized the fermion multiplets while
keeping all the symmetry-breaking and light Higgs fields at one point.
We show for relatively small moose, one can
arrange for sufficient overlaps via higher-dimension operators involving
the link fields to account for fermion masses and inter-generational mixing.
For larger moose, we assume instead that the unified boundary condition on
the gauge couplings is simply present (or alternately, the $Z_3$ symmetry
is an accidental symmetry of only the gauge and Higgs sectors) so that
matter can be embedded asymmetrically.  Then the overlap between matter fields
and light Higgs fields can be made maximal and no restriction follows on the
size of the moose.  Alternatively, such models provide the freedom to
adjust matter-matter and matter-Higgs overlaps in a way that can account in
part for the flavor hierarchies of the standard model.  This is one
advantage of the models introduced here over minimal trinified scenarios.

Aside from demonstrating how one may deconstruct the 5D trinified
theory described in Ref.~\cite{ccjc}, we have arrived at at class of
4D theories that are interesting in their own right.  This is not a
surprise, as the same is true of deconstructed 5D theories based on
SU(5) symmetry that have been studied in the literature~\cite{ckt}.
The present work therefore provides a new framework for future study
of a range of familiar issues, including the implementation of
solutions to the doublet-triplet splitting problem, proton decay
phenomenology, the origin of flavor via combined unified and
horizontal symmetries, construction of explicit symmetry breaking
sectors, {\em etc}.  Models based on the idea of nonsupersymmetric
triplicated trinification~\cite{trin5} may also be worthy of study in
this context.


\begin{acknowledgments}
CDC thanks Joshua Erlich and Marc Sher for useful comments, and the
NSF for support under Grant No.~PHY-0140012.
\end{acknowledgments}

\end{document}